\documentclass[sn-mathphys-num]{sn-jnl}

\usepackage{graphicx}%
\usepackage{multirow}%
\usepackage{amsmath,amssymb,amsfonts}%
\usepackage{amsthm}%
\usepackage{mathrsfs}%
\usepackage[title]{appendix}%
\usepackage{xcolor}%
\usepackage{textcomp}%
\usepackage{manyfoot}%
\usepackage{booktabs}%
\usepackage{algorithm}%
\usepackage{algorithmicx}%
\usepackage{algpseudocode}%
\usepackage{listings}%

\theoremstyle{thmstyleone}%
\theoremstyle{thmstyletwo}%

\theoremstyle{thmstylethree}%

\begin{document}

\title[Article Title]{Primordial Nucleosynthesis with Non-Extensive Statistics}


\author*[1]{\fnm{C. A.} \sur{Bertulani}}\email{carlos.bertulani@tamuc.edu }

\author[2]{\fnm{} \sur{Shubhchintak}}\email{shubhchintak\_phy@pbi.ac.in}

\affil*[1]{\orgdiv{Department of Physics and Astronomy}, \orgname{Texas A\&M University-Commerce}, \orgaddress{\city{Commerce}, \postcode{75429}, \state{Texas}, \country{USA}}}

\affil[2]{\orgdiv{Department of Physics}, \orgname{Punjabi University, Patiala}, \orgaddress{\city{Patiala}, \postcode{147002}, \state{Punjab}, \country{India}}}


\abstract{The conventional Big Bang model successfully anticipates the initial abundances of $^2$H(D), $^3$He, and $^4$He, aligning remarkably well with observational data. However, a persistent challenge arises in the case of $^7$Li, where the predicted abundance exceeds observations by a factor of approximately three. Despite numerous efforts employing traditional nuclear physics to address this incongruity over the years, the enigma surrounding the lithium anomaly endures. In this context, we embark on an exploration of Big Bang nucleosynthesis (BBN) of light element abundances  with the application of Tsallis non-extensive statistics. A comparison is made between the outcomes obtained by varying the non-extensive parameter $q$ away from its unity value and both observational data and abundance predictions derived from the conventional big bang model. A good agreement is found for the abundances of $^4$He, $^3$He and $^7$Li, implying that the lithium abundance puzzle might be due to a subtle  fine-tuning of the physics ingredients used to determine the BBN. However, the deuterium abundance deviates from observations.}




\maketitle

\section{Big Bang Nucleosynthesis}\label{sec1}

While the cosmological big bang model aligns with numerous observations crucial for our comprehension of the universe, the comparison between model-based calculations and actual observations is not straightforward. This complexity arises due to poorly understood time-sensitive effects and systematic errors affecting the data. Nevertheless, the big bang model stands as the primary tool for investigating the physics of the early universe within the time frame of $3-20$ minutes. Beyond this point, the decreasing temperature and density prevent nuclear fusion, hindering the formation of elements heavier than beryllium \cite{ABG48,Alp50,Alp53,WFH67,Zel64,Peeb09,Peeb19}.

The model corresponds with the observed cosmic microwave background (CMB) radiation temperature of 2.275 K \cite{Not11}, offering insights into various scientific realms, including nuclear and particle physics. Additionally, calculations based on the big bang model remain consistent with the number of light neutrino families, $N_\nu =3$. The literature on the subject suggests a range for $N_\nu$ between $1.8$ and $3.9$ (see, for instance, Ref. \cite{Oli02}). Notably, measurements from experiments at CERN indicate $N_\nu = 2.9840 \pm 0.0082$ based on the $Z_0$ width \cite{PDG20}.

In the framework of the big bang model, nearly all neutrons ultimately combine to form $^4$He, introducing a dependence of the abundance of $^4$He on $N_\nu$ and the neutron lifetime $\tau_n$. The dependence on $\tau_n$ influences Big Bang Nucleosynthesis (BBN) in two distinct ways. Firstly $\tau_n$ impacts weak reaction rates due to its connection to the weak coupling constant. A longer (or shorter)  $\tau_n$ implies that the reaction rates remain lower (or higher)  than the Hubble expansion rate until a higher (or lower)  freeze-out temperature, significantly affecting the neutron-to-proton equilibrium ratio at freeze-out.

The ratio of neutrons to protons in thermal equilibrium, denoted as $n/p$, is approximately characterized by $n/p = \exp[-\Delta m/k T] \approx 1/6$, where  $T$ stands for the temperature at the weak freeze-out, and $\Delta m\sim 1.3$ MeV signifies the mass difference of neutron and proton. Another significant effect of $\tau_n$ arises from neutron decay occurring between the weak freeze-out ($t\sim 1$ s) and the onset of nucleosynthesis ($t\sim 200$ s), causing a decrease in the $n/p$ ratio to approximately $1/7$. A longer $\tau_n$ corresponds to a higher predicted BBN helium abundance. Here, we adhere to the most recent experimental determination of $\tau_n$, specifically $\tau_n=877.75 \pm 0.28_{stat} + 0.22/-0.16_{syst}$ s, as documented in recent experiments \cite{Gonz21} (for an extensive review of neutron lifetime, see Ref. \cite{FG11}). The ramifications of variations in $\tau_n$ on BBN predictions have been recently explored in Ref. \cite{yeh23}.

The baryonic density in the universe, derived from astronomical observations anisotropies in the CMB radiation, places constraints on the value of the baryons to photon ratio, denoted as $\eta$. This ratio remains constant as the universe expands. Calculations within the BBN model align with the value determined from WMAP observations, indicating $\eta = 6.104 \pm 0.055 \times 10^{-10}$ \cite{Agh2020}.

Our focus in this brief review centers on the abundances of light elements. During the initial stages of the universe, specifically within the first 20 minutes, when conditions were sufficiently dense and hot to facilitate nuclear reactions, the temperature of the primordial plasma underwent a substantial decrease from a few million electronvolts (MeV) to approximately 10 kiloelectronvolts (keV). This environment gave rise to the production of light nuclides such as $^2$H, $^3$He, $^4$He, and, to a lesser extent, $^7$Li, through a complex network of nuclear reactions. The final abundances for these nuclides can be determined through various observational techniques, spanning diverse astrophysical environments.

However, discrepancies in lithium (Li) abundances observed in metal-poor stars, compared to inferences from WMAP data, have sparked extensive inquiries into both BBN and the stellar mixing processes that potentially influence Li abundance. The findings of these investigations remain inconclusive, emphasizing the need for further studies on the abundances of light elements in low-metallicity stars and in extragalactic HII regions, i.e., denser, collapsed regions in giant molecular clouds where stars are forming. Additionally, refining estimates using BBN becomes crucial to address this issue comprehensively. High-resolution spectroscopic analyses of stellar and interstellar matter need to be integrated with  nuclear physics experiments and nucleosynthesis models and nuclear theories to gain a deeper understanding of these intricacies \cite{Fie11,cyburt16}.

The calculation of nuclear reaction rates during BBN relies on fundamental inputs, one of which is the Maxwell-Boltzmann distribution describing the kinetic energy of particles in a plasma. This distribution is grounded on basic assumptions of the Boltzmann-Gibbs statistics, namely,  (a) the particle collision time is significantly shorter than the average time between collisions, (b) interactions are localized, (c) velocities of two particles are uncorrelated, and (d) there is a minimal energy transfer  to and from collective variables and therefore energy of the colliding particles is locally conserved. When conditions (a) and (b) are not satisfied, the effective two-body interaction becomes non-local, dependent on the energy and momentum of the ions. Assuming that the one-particle energy distribution follows the Maxwellian pattern, additional assumptions on particle correlations are needed to prove that the relative-velocity distribution is still Maxwellian.

While the Boltzmann-Gibbs (BG) model of statistical mechanics holds well in numerous situations, recent theoretical efforts have increasingly explored alternative approaches, treating BG statistics as a limit within more general theories \cite{Ts88} (see also, \cite{Ren60}). These alternative theories aim to describe systems with long-range interactions and memory effects, departing from the assumptions of ergodicity. One prominent alternative to BG statistics, as suggested by C. Tsallis \cite{Ts88,GT04}. This approach, known as {\it non-extensive statistics}, has gained popularity in recent years (for in-depth details, refer to comprehensive reviews \cite{GT04,TGS05,Ts09}). While traditional statistical mechanics assumes energy to be an ``extensive" variable, proportional to the system size, and entropy to be extensive, this justification may be valid for short-range interactions binding matter. However, when dealing with long-range interactions, particularly in the context of gravity, entropy is found to be non-extensive \cite{FL01,Lim02,TS03,TS04,CS05}.

In traditional statistics, determining the mean values of system quantities like energy, the number of molecules, and volume involves seeking the probability distribution that maximizes entropy while satisfying constraints for the correct average values of those quantities. Cognizant of this, Tsallis introduced a novel, non-extensive entropy, commonly known as the {\it Tsallis entropy}, and advocated for its maximization under the same constraints, replacing the conventional (BG) entropy. A family of Tsallis entropies are characterized by a real-valued parameter $q$ that quantifies the departure from extensivity, reverting to the usual definition of entropy when $q = 1$. One has demonstrated in various scenarios that traditional results from statistical mechanics can be translated into this new framework \cite{Ts09}. These entropies,  result in probabilities following power laws instead of the typical exponential laws found in the standard statistics \cite{Ts09}. In many instances where the Tsallis formalism is employed, such as in Ref. \cite{Pes01}, the non-extensive parameter is held constant $q \simeq 1$, i.e., close to the value yielding conventional statistical mechanics. Some studies have explored substantial deviations of $q$ from unity to elucidate various phenomena across diverse scientific domains \cite{GT04}.

We will demonstrate that the foundational Maxwell-Boltzmann (MB) distribution, crucial to big bang and stellar evolution nucleosynthesis, undergoes significant modification under non-extensive statistics when $q$ deviates substantially from unity. Consequently, this alteration exerts a profound impact on the predictions of BBN. A robust prior justification for a substantial deviation of $q$ from unity in the BBN is lacking. Specifically, since matter is presumed to be in equilibrium with radiation during BBN, any modification to the MB distribution of velocities would influence the Planck distribution of photons. Analyses of temperature fluctuations in the CMB have demonstrated that a Planck distribution modified by the Tsallis statistics accurately describes temperature fluctuations in the CMB as observed with WMAP and $q = 1.045 \pm 0.005$, a value close to unity but not precisely equal \cite{Ber07}. Notably, temperature distributions based on Boltzmann-Gibbs statistics,  inadequately describe CMB temperature fluctuations \cite{Ber07}. These fluctuations, possibly allow for even larger variations in $q$ and might manifest during the BBN, consequently influencing the exponentially small tail of the velocity distribution.

Building on the accomplishments of the BBN model, it is reasonable to trust that it can impose stringent constraints on the parameter $q$ of the non-extensive statistics. Within the scientific literature, endeavors to address the lithium problem often involve exploring various facets of ``new physics" \cite{Fie11}. This study contributes to the array of novel approaches, with implications extending far beyond the expected scope for resolving the lithium problem within BBN \cite{BFH13,Hou17,Oliv16}. If Tsallis statistics effectively characterizes deviations in the tails of statistical distributions, then BBN becomes a valuable tool for probing such deviations.  Indeed, the potential for a departure from the Maxwellian distribution and the repercussions of modifying the MB distribution's tail for nuclear fusion in stars have been explored in previous studies \cite{MQ05,HK08,Deg98,Cor99} and in recent studies it has been shown that a small departure of the parameter $q$ from the unity is enough to solve the lithium puzzle \cite{Hou17}.

\section{Extensive versus Non-Extensive Statistics}

\subsection{Extensive Statistics}
Consider two levels separated by an energy $E$ with $n_0$ particles occupying the lowest energy level and $n_1$ the higher energy one. For identical particles, the number of ways that one can arrange the system is $\Omega = N!/n_0! n_1!$, where $N=n_0+n_1$. The number $\Omega$ attains very large values for macroscopic systems and a better measure of the number of possibilities that the system can rearrange is  $S=k \ln \Omega$. This is the usual definition of entropy and $k$ is identified as the Boltzmann constant. For the aforementioned two-level problem we get $S=k[\ln N! - \ln n_0! - \ln n_1!]$. If an additional amount $E$ of energy is added to the system, one particle is promoted to the higher level and the entropy increases to $S'$ with the replacements $n_0 \rightarrow n_0 - 1$ and  $n_1 \rightarrow n_1 + 1$. The additional entropy gain is $\Delta S = S'-S=k\ln (n_0/n_1)$, for $n_1 \gg 1$. If we compare to the thermodynamics relation for the change in entropy, we have $\Delta S = \Delta E/T=E/T$, which implies that $n_1=n_0 \exp(-E/kT)$, the famous Boltzmann law of statistics.

Extending this logic to multiple energy levels, one defines occupation probabilities proportional to particle numbers in each level $i$. Normalizing these probabilities, yields
\begin{equation}
P(E_i) = \frac{1}{\cal Z} \exp\left(- \frac{E_i}{kT}\right), \ \ \ \ {\rm where} \ \ \  {\cal Z} = \sum_i P(E_i),
\end{equation}
for the occupation probability of the state $i$ with energy $E_i$. ${\cal Z}$ is know as the partition function. 

The Boltzmann law is a cornerstone of thermodynamics and statistics. For non-interacting particles in thermal equilibrium, one can define $f(v)dv$ as the probability of encountering a particle with velocity $v$. This leads to the Maxwell-Boltzmann (MB) distribution:
\begin{equation}
    f(v) = 4\pi v^2 \left( \frac{m}{2\pi kT}\right)^{3/2} \exp\left( -\frac{mv^2}{2kT}\right) ,
\end{equation}
which we focus on in this review.

The property of extensivity can be easily understood if the Boltzmann-Gibbs entropy is expressed as
\begin{equation}
S_{BG}=-k \sum_i p_i \ln p_i , \label{sbg}
\end{equation}
where $p_i$ represents the probability of finding the system in the $i$-th microstate, and $\sum_i p_i = 1$. Considering two independent systems labeled $A$ and $B$, the combined probability of finding a state $i+j$ in a  microstate $i$ of $A$ and $j$  of $B$, is:
\begin{equation}
p^{A+B}_{i+j} = p^A_i \cdot p^B_j.
\end{equation}
Inserting this into Eq. \ref{sbg}, implies that the Boltzmann-Gibbs entropy adheres to the relationship
\begin{equation}
S_{A+B}=S_A +S_B.
\end{equation}
Hence, entropy derived from Boltzmann-Gibbs statistics is recognized as an \textit{extensive} quantity since, akin to volumes of thermodynamic systems, it adds extensively. 

\subsection{Non-Extensive Statistics}
Boltzmann-Gibbs statistics provides a sound framework for treating ergodic systems, where given sufficient time, any state has a non-zero probability $p_i$ of occurrence. However, particle correlations and the breakdown of ergodicity often push systems into out-of-equilibrium states, rendering Boltzmann-Gibbs statistics inadequate. Furthermore, some assumptions inherent to Boltzmann-Gibbs statistics, particularly when applied to interacting systems, prove overly restrictive. They are not universally valid, even in systems of particles at thermodynamic equilibrium. Indeed, alternatives to Boltzmann-Gibbs statistics are available and frequently utilized in the literature~\cite{Ren60,Ts88,GT04}.

Within the realm of non-extensive statistics \cite{Ts88}, a departure from tradition emerges as the conventional entropy yields to an alternative formulation, due to Tsallis~\cite{Ts88}:
\begin{equation}
{\cal S}_q=k \frac{1-\sum_i p^q_i} {q-1},
\end{equation}
where $q$ denotes a real number. When $q=1$, ${\cal S}_q= {\cal S}_{BG}$, illustrating Tsallis statistics as a smooth extension of the Boltzmann-Gibbs entropy.

Following that, it can be easily shown that
\begin{equation}
{\cal S}_q(A+B)={\cal S}_q(A) +{\cal S}_q(B)+\frac{(1-q)}{k}{\cal S}_q(A){\cal S}_q(B).
\end{equation}
Here, the parameter $q$ acts as a gauge for \textit{non-extensivity}, a concept pioneered by Tsallis in the development of statistical mechanics grounded in this generalized entropy \cite{Ts09}.

A significant implication of this non-extensive framework is the deviation of the distribution function maximizing $S_q$ from the Maxwellian \cite{Sil98,Lim00,Mu06}. With $q=1$, the MB distribution is faithfully replicated. However, for $q<1$, high-energy velocities become notably more probable compared to the extensive distribution. Conversely, when $q>1$, high-energy velocities become less probable, with a subsequent cutoff beyond which no velocities are possible.

\subsection{Nuclear Reaction Rates with Extensive Statistics}

Within stellar environments, the rate of thermonuclear reactions, weighted with a MD in the plasma, is given by \cite{Fow67}:
\begin{eqnarray}
R_{ij}&=&\frac{N_{i}N_{j}}{1+\delta_{ij}} \langle \sigma v \rangle \nonumber \\&=& \frac{N_{i}N_{j}}{1+\delta_{ij}}\left(\frac{8}{\pi\mu}\right)^{\frac{1}{2}}\left(\frac{1}{kT}\right)^{\frac{3}{2}}
\int_{0}^{\infty}dE\,S(E) \exp\left[-\left(\frac{E}{kT}+2\pi\eta(E)\right)\right]. \label{rij}
\end{eqnarray}
Here,  $T$ denotes the temperature, $S(E)$ stands for the astrophysical S-factor, $\sigma$ denotes the nuclear fusion cross section, $v$ signifies the relative velocity between the $ij$-pair, $E=\mu v^2/2$ denotes their relative motion energy, $\mu$ stands for their reduced mass, $N_i$ represents the number of nuclear species $i$,   and $\eta=Z_iZ_je^2/\hbar v$ is the Sommerfeld factor, where $Z_i$ denotes the charge of the $i$-th particle. The factor $1+\delta_{ij}$ in the denominator avoids double-counting when $i=j$. 
A modified expression for the reaction rates, namely, $N_A \langle \sigma v \rangle$,  can be expressed in units of cm$^3$ mol$^{-1}$ s$^{-1}$, where $N_A$ denotes Avogadro's number, and $\langle \sigma v \rangle$ is given in  Eq. \eqref{rij} with the substitution of the Maxwell distribution $f(E)$ by Eq. \eqref{fene} defined below. 

To factor out the steep energy dependence of reaction cross sections on energy, one usually defines:
\begin{equation}
\sigma(E) = \frac{S(E)}{E}\exp\left[-{2\pi\eta(E)}\right].
\label{se}
\end{equation}
We write $2\pi \eta=b/\sqrt{E}$, with
\begin{equation}
b=0.9898 Z_iZ_j\sqrt{A} \ {\rm MeV}^{1/2}, \label{b}
\end{equation}
where $A$ represents the reduced mass in atomic mass units. The S-factor exhibits a weaker dependence on $E$, unless it approaches  a resonance, where it manifests a strong peak. For a weak $S(E)$ energy dependence, the reaction rate integral in Eq. \ref{rij} is peaked at the ``Gamow energy"  
\begin{equation}
E_0=0.122(Z_i^2Z^2_jA)^{1/3}T_9^{2/3} \ {\rm MeV}, \label{gamow}
\end{equation}
and the ``Gamow width" of the peak is 
\begin{equation}
\Delta E=0.2368(Z_i^2Z^2_jA)^{1/6}T_9^{5/6} \ {\rm MeV}, \label{gamoww}
\end{equation}
where $T_9$ it the temperature in units of $10^9$ K, and $A$ denotes the reduced mass in atomic mass units (amu). Regarding charged particle reactions in the BBN, a high precision (within 0.1\%) is attained by employing integration limits ranging from $E_0 - 5\Delta E$ to $E_0+5\Delta E$, where $\Delta E$ is determined by Eq. (\ref{gamoww}). 

Neutron-induced reactions are a subset of paramount importance for BBN, particularly including the p(n,$\gamma$)d, ${^3}$He(n,p)t, and $^7$Be(n,p)$^7$Li reactions. The cross section at low energies typically scales proportionally to $1/v$, where the neutron velocity is represented by $v=\sqrt{2mE}/\hbar$. Consequently, it's sometimes advantageous to reformulate Eq. \eqref{se} as
\begin{equation}
\sigma(E) = \frac{S(E)}{E}= \frac{R(E)}{\sqrt{E}},
\label{sen}
\end{equation}
where $R(E)$ is a smooth function of energy akin to an $S$-factor.

\section{Nuclear Reaction Rates}

\subsection{Nuclear Reaction Rates with Non-Extensive Statistics}
The non-extensive characterization of the Maxwell-Boltzmann distribution is realized through the replacement $f(E) \rightarrow f_q(E)$, as detailed by \cite{Ts09}:
\begin{equation}
f_q(E)=\left[1-\frac{q-1}{kT}E\right]^{\frac{1}{q-1}} \ \ \stackrel{ q\rightarrow1}{\longrightarrow}\ \ \exp\left(-\frac{E}{kT}\right),
\quad 0<E<\infty. \label{fene}
\end{equation}
If $q-1<0$, this equation yields real values for any value of $E\ge 0$. On the other hand, if $q-1>0$, it is clear that $f(E)$ is real only if the expression within square brackets remains positive. In other words,
\begin{equation}
0 \le E \le \frac{kT}{q-1}, \  {\rm for} \ \ q\ge 1,\ \ \ \ \ {\rm and} \ \ \ \ \
0\le E, \  {\rm for} \ \ q\le 1.
\end{equation}
Consequently, for $0<q<1$, the range is $0<E<\infty$, while if $1<q<\infty$, it becomes $0<E<E_{\mathrm{max}}={kT}/(q-1)$.

Using this alternate statistics, the nuclear reaction rates transform into:
\begin{equation}
R_{ij}= \frac{N_{i}N_{j}}{1+\delta_{ij}} \int_{0}^{E_{max}}dE\,S(E){\cal M}_{q}(E,T),\label{Iq}
\end{equation}
with the ``modified" Gamow energy distribution given by
\begin{eqnarray}
{\cal M}_{q}(E,T)&=&{\cal A}(q,T)\left(1-\frac{q-1}{kT}E\right)^{\frac{1}{q-1}}e^{-b/\sqrt{{E}}} \nonumber \\
&=&{\cal A}(q,T) \left(1-\frac{q-1}{0.08617T_9}E\right)^{\frac{1}{q-1}}\exp\left[-0.9898 Z_iZ_j\sqrt{\frac{A}{E}}\right],
\label{MqET}\end{eqnarray}
representing the non-extensive counterpart of the nuclear astrophysics input in the BBN, where $E_{max}=\infty$ if $0<q<1$, and $E_{\mathrm{max}}={kT}/(1-q)$ if $1<q<\infty$, and with $E$ given in MeV units. The function ${\cal A} (q,T)$ is a normalization constant dependent on the temperature $T$ and on the non-extensive parameter $q$.

In the case of neutron induced reactions, the modified Gamow distribution function in Eq. \eqref{Iq} can be expressed as
\begin{equation}
{\cal M}_{q}(E,T)={\cal A} (q,T)f_q(E)={\cal A} (q,T)\left(1-\frac{q-1}{kT}E\right)^{\frac{1}{q-1}}. 
\label{MqET2}
\end{equation}
The omission of the tunneling factor in this equation decreases the reliance of the reaction rates on the non-extensive parameter $q$.

The quest for a unified statistical framework encompassing Fermi-Dirac, Bose-Einstein, and Tsallis statistics for particles in a plasma has been a focal point of research, as discussed in Ref. \cite{Buy00}. In this pursuit, a non-extensive statistical distribution for both fermions and bosons was proposed, represented by the expression:
\begin{equation}
n_q^{\pm} (E)=  \frac{1}{\left[1-(q-1)\frac{(E-\mu)}{kT}\right]^{\frac{1}{q-1}}\pm 1}, \label{nnon}
\end{equation}
where $\mu$ denotes the chemical potential. This formulation seamlessly transitions to the Fermi distribution ($n^+$) as $q$ approaches 1 and to the Bose-Einstein distribution ($n^-$) for photons when $\mu=0$ and $q$ tends to 1.
Furthermore, by incorporating the appropriate modifications, Planck's law governing the distribution of radiation is derived by augmenting $n^-$ in the equation above (with $\mu=0$) with $\hbar \omega^2/(4\pi^2 c^2)$, where $E=\hbar \omega$.
For electrons, the number density can be derived from $n^+$ in the equation above, with considerations for relativistic or non-relativistic regimes, necessitating the inclusion of proper phase factors. Additionally, normalization factors ${\cal A}^\pm (q,T)$ are introduced to ensure consistency, as previously established in the literature.

During the early universe, electron density experiences significant variation with temperature, with electron number densities reaching approximately $10^{32}$/cm$^3$ at $T_9=10$, vastly surpassing the electron density at the sun's core, approximately $10^{26}$/cm$^3$. This heightened electron density stems from the copious production of $e^+e^-$ pairs by abundant photons during BBN. However, despite the substantial electron densities, nuclear reactions remain largely unaffected, with the enhancement of nuclear reaction rates due to electron screening proving negligible, as demonstrated by reference \cite{WBB11}.

\begin{center}
\begin{figure}[t]
\centering
\includegraphics[width=85mm]{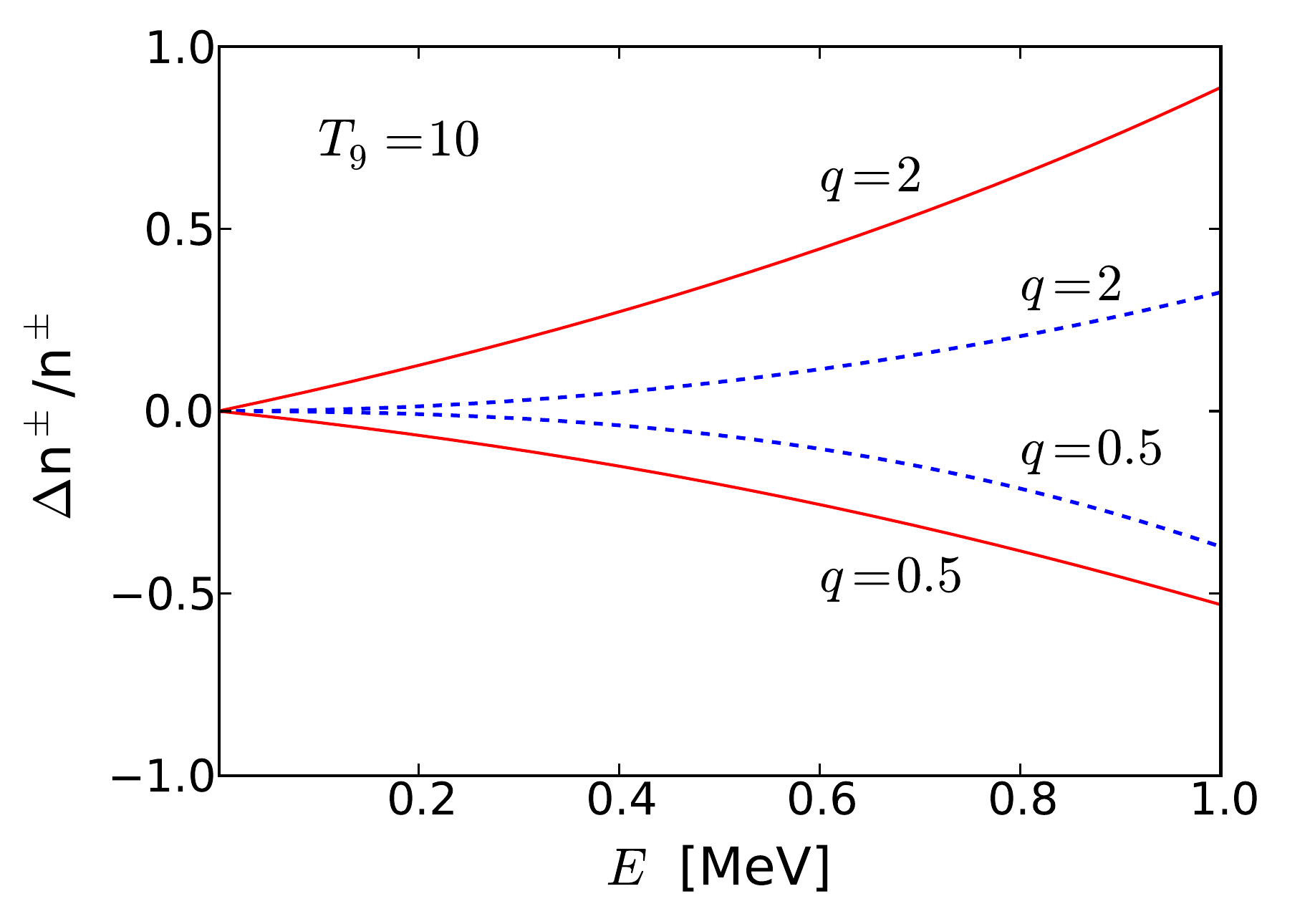}
\caption{The difference between non-extensive $n_q$ and extensive $n=n_{q\rightarrow 1}$ statistics expressed in terms of $(n_q^{\pm}-n^\pm)/n^\pm$ for Fermi-Dirac statistics $n^+$ (solid lines) and for Bose-Einstein statistics $n^-$ (dashed lines).  In both cases,  $\mu=0$ is used for $q=2$, $q=0.5$, and $T_9=10$. }\label{fdn} 
\end{figure}
\end{center}

The electron Fermi energy at these densities remains much smaller than $kT$ for energies relevant to BBN, permitting the utilization of $\mu=0$ in Eq. \eqref{nnon} for $n^+$. In Figure \ref{fdn}, the relative difference ratio $(n_q^{\pm}-n^\pm)/n^\pm$ between non-extensive ($n_q$) and extensive ($n=n_{q\rightarrow 1}$) statistics is illustrated. Solid curves represent Fermi-Dirac (FD) statistics ($n^+$), while dashed curves depict Bose-Einstein (BE) statistics ($n^-$), with results shown for $q=2$ and $q=0.5$, at $T_9=10$. Notably, non-extensive distributions are amplified for $q>1$ and suppressed for $q<1$ compared to respective FD and BE quantum distributions. These deviations grow larger with energy, with the non-extensive electron distribution roughly twice as large as the usual FD distribution at $E_e = 1$ MeV, at $T_9=10$.

Although numerical results with modified FD and BE distributions are not presented here, it is anticipated that these generalizations will profoundly impact the freezout temperature and the neutron to proton (n/p) ratio. The freezout temperature, marking when the weak reaction rate, $\Gamma \sim \left<\sigma v\right>$, for the weak reaction $\nu_e + n \rightarrow p+e^-$ becomes slower than the universe's expansion rate, remains unaffected by non-extensive statistics due to low particle densities compared to $kT$, allowing $\mu$ to be set to zero. The adoption of non-extensive quantum distributions, as in Eq. \eqref{nnon}, yields the same temperature powers as those predicted by the FD and BE distributions, ensuring consistency in reaction rates.

\subsection{Non-Maxwellian Distribution for Relative Velocities}

It's important to highlight that while a Maxwellian distribution governs individual particle energies, it doesn't inherently ensure a Maxwellian distribution of relative velocities. To achieve this, additional assumptions regarding particle correlations are necessary to establish the distribution of relative velocities, which is pivotal for accurate rate calculations. This matter has been extensively studied in references \cite{MQ05,Ts88,Kan05}, where it's demonstrated that non-Maxwellian distributions arise from non-extensive statistics.

In this context, we illustrate that if the particle velocity distribution deviates from Maxwellian behavior as described by Eq. \eqref{fene}, then an adjustment can be made to the two-particle relative velocity to accommodate the recoil of the center-of-mass. Denoting the kinetic energy of a particle as $E_i$, this distribution is represented by:
\begin{equation}
f_q^{(i)} = \left(1 - \frac{q - 1}{kT}E_i\right)^{\frac{1}{q - 1}} \ \ \stackrel{ q\rightarrow1}{\longrightarrow} \exp{\left[-\left(\frac{E_i}{kT}\right)\right]}.\label{onepart}
\end{equation}

Let's consider the equation:
\begin{equation}
f_q^{(i)} = \left(1 - \frac{q - 1}{kT}E_i\right)^{\frac{1}{q - 1}} \nonumber
\end{equation}
Expressing the two-particle energy distribution as $f_q^{(12)}=f_q^{(1)}f_q^{(2)}$, and exponentiating the Tsallis distribution, we obtain:
\begin{equation}
f_q^{(i)} = \exp\left\{\frac{1}{q - 1} \left[\ln\left(1 - \frac{q - 1}{kT} E_i\right)\right]\right\}.
\end{equation}
This leads to the simplification of the product $f_q^{(1)}f_q^{(2)}$ as:
\begin{equation}
f_q^{(12)} = \exp\left\{\frac{1}{q - 1} \left[\ln\left(1 - \frac{q - 1}{kT} E_1\right)\ln\left(1 - \frac{q - 1}{kT}E_2\right)\right]\right\}.
\end{equation}

Considering $E_i = {m_i v_i^2}/{2}$, where $E_1 + E_2 = {\mu v^2}/{2} + {M V^2}/{2}$, with $\mu$ representing the reduced mass of the two particles, $M = m_1 + m_2$, $v$ being the relative velocity, and $V$ as the center-of-mass velocity, the product inside the natural logarithm simplifies to:
\begin{eqnarray}
&& 1 - \frac{1 - q}{kT} \left(\frac{\mu v^2}{2} + \frac{M V^2}{2}\right) + \left(\frac{1 - q}{kT}\right)^2 \frac{\mu v^2}{2}\frac{M V^2}{2} \nonumber \\
&&=\left(1 - \frac{1 - q}{kT} \frac{\mu v^2}{2}\right)\left(1 - \frac{1 - q}{kT} \frac{M V^2}{2}\right)\label{factoriz}
\end{eqnarray}

Therefore, the two-body distribution can be factorized into a product of relative and center-of-mass components:
\begin{eqnarray}
f_q^{(12)}(v,V,T) = f_q^{(rel)}(v,T) f_q^{(cm)}(V,T),
\end{eqnarray}
where
\begin{eqnarray}
f_q^{(rel)}(v,T)&=&{\cal A}_{rel}(q,T)\left(1 - \frac{1 - q}{kT} \frac{\mu v^2}{2}\right)^{\frac{1}{q - 1}},\nonumber\\
f_q^{(cm)}(V,T)&=&{\cal A}_{cm}(q,T)\left(1 - \frac{1 - q}{kT} \frac{M V^2}{2}\right)^{\frac{1}{q - 1}}.
\end{eqnarray}

The normalization constants are determined by the condition
\begin{equation}
\int d^3v d^3V\,f_q^{(12)}(v,V,T) = 1.
\end{equation}
Given the factorization of the distribution, unit normalization can be obtained for the relative and center-of-mass distributions separately. Hence, the distribution required in the reaction rate formula is:
\begin{equation}
f_q(v,T) = \int d^3V f_q^{(12)}(v,V,T)= f_q^{(rel)}(v,T),
\end{equation}
which adopts the same form as the absolute distribution.

In the limit $q \rightarrow 1$, the two-particle distribution reduces to a Gaussian, where the last term in the left-hand side of Eq. \eqref{factoriz} drops out:
\begin{equation}
f_q^{(12)}= {\cal A}(q,T) \exp\left[-\left(\frac{{\mu v^2}/{2} + {MV^2}/{2}}{kT}\right)\right],
\end{equation}
as anticipated.

\section{Fixing the BBN Lithium Problem}

\subsection{Non-Extensive Energy Distributions}
\begin{center}
\begin{figure}[t]
\centering
\includegraphics[width=0.7\textwidth]{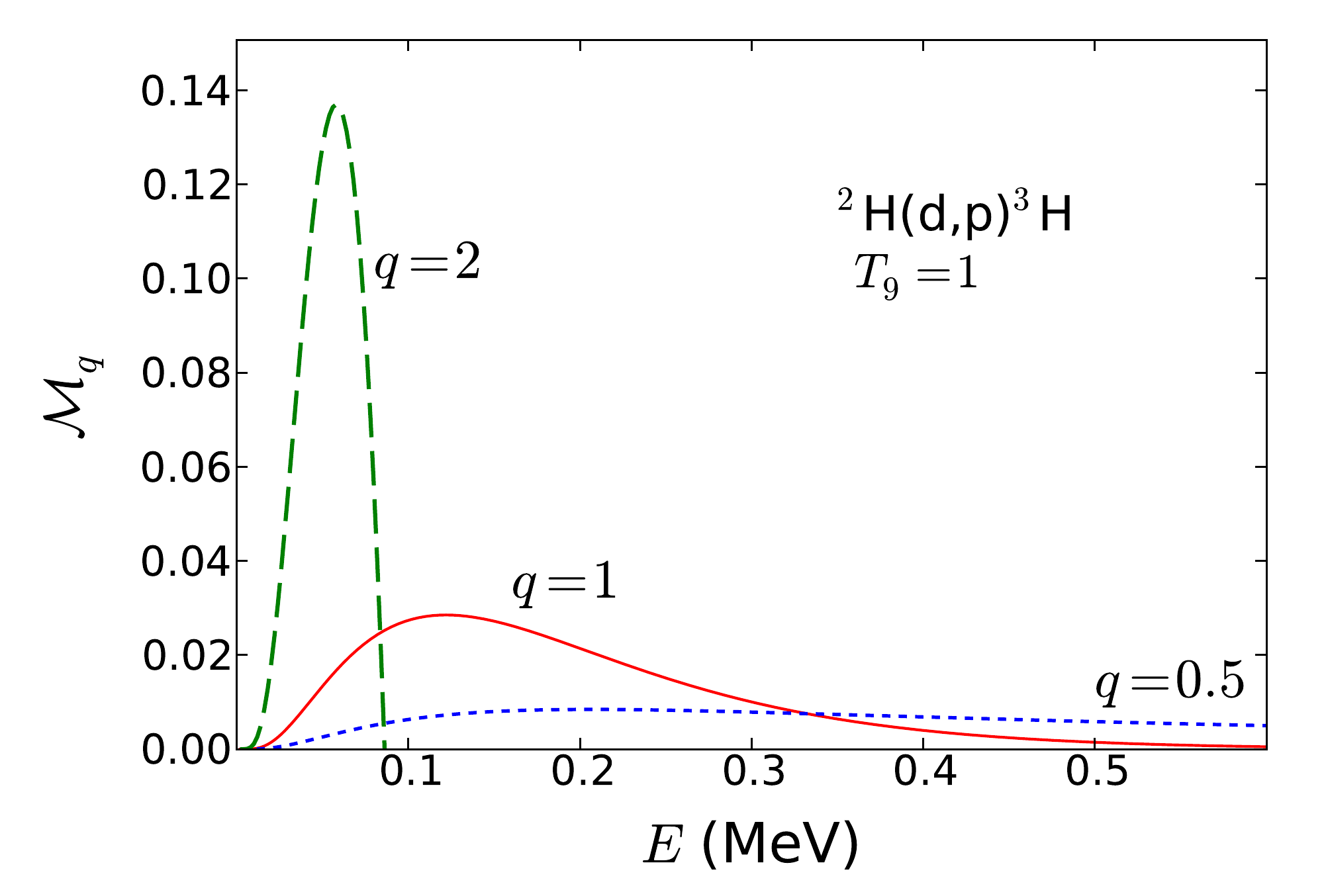}
\caption{Non-extensive Gamow distributions ${\cal M}_{q}(E,T)$ of deuterons of importance for the reaction $^2$H(d,p)$^3$H at $T_9=1$. The solid line, representing $q=1$, refers to the use of a  MB distribution. Also depicted are the outcomes obtained if one employs non-extensive MB distributions. For the sake of understanding the deviation from the usual MB distribution, we exploit very large deviations for $q$ from the unity, namely, $q=0.5$ (dotted line) and $q=2$ (dashed line).}\label{fmaxwell}
\end{figure}
\end{center}

Referring to abundant literature on non-extensive statistics (see, for example, \cite{MQ05,HK08,Deg98,Cor99,Ts88}), it is commonly expected that the non-extensive parameter $q$ remains close to unity. Nonetheless, to explore the ramifications of non-Maxwellian distributions on BBN, we investigate scenarios where $q$ significantly deviates from unity. This investigation aims to deepen our understanding of the underlying physics departing from Boltzmann-Gibbs (BG) statistics. Figure \ref{fmaxwell} illustrates Gamow energy distributions of deuterons of importance for the reaction $^2$H(d,p)$^3$H at $T_9=1$.  The solid line, representing $q=1$, refers to the use of a  MB distribution. Also depicted are the outcomes obtained if one employs non-extensive MB distributions. For the sake of understanding the deviation from the usual MB distribution, we exploit very large deviations for $q$ from the unity, namely, $q=0.5$ (dotted line) and $q=2$ (dashed line).   It is noteworthy that  higher kinetic energies become more accessible when $q<1$ compared to the extensive case $q=1$, whereas high energies are less probable for $q>1$, with a distinct kinetic energy cutoff. Here, for $q=2$, the cutoff occurs at $0.086$ MeV, equivalent to $86$ keV.

\begin{center}
\begin{figure}[t]
\centering
\includegraphics[width=0.7\textwidth]{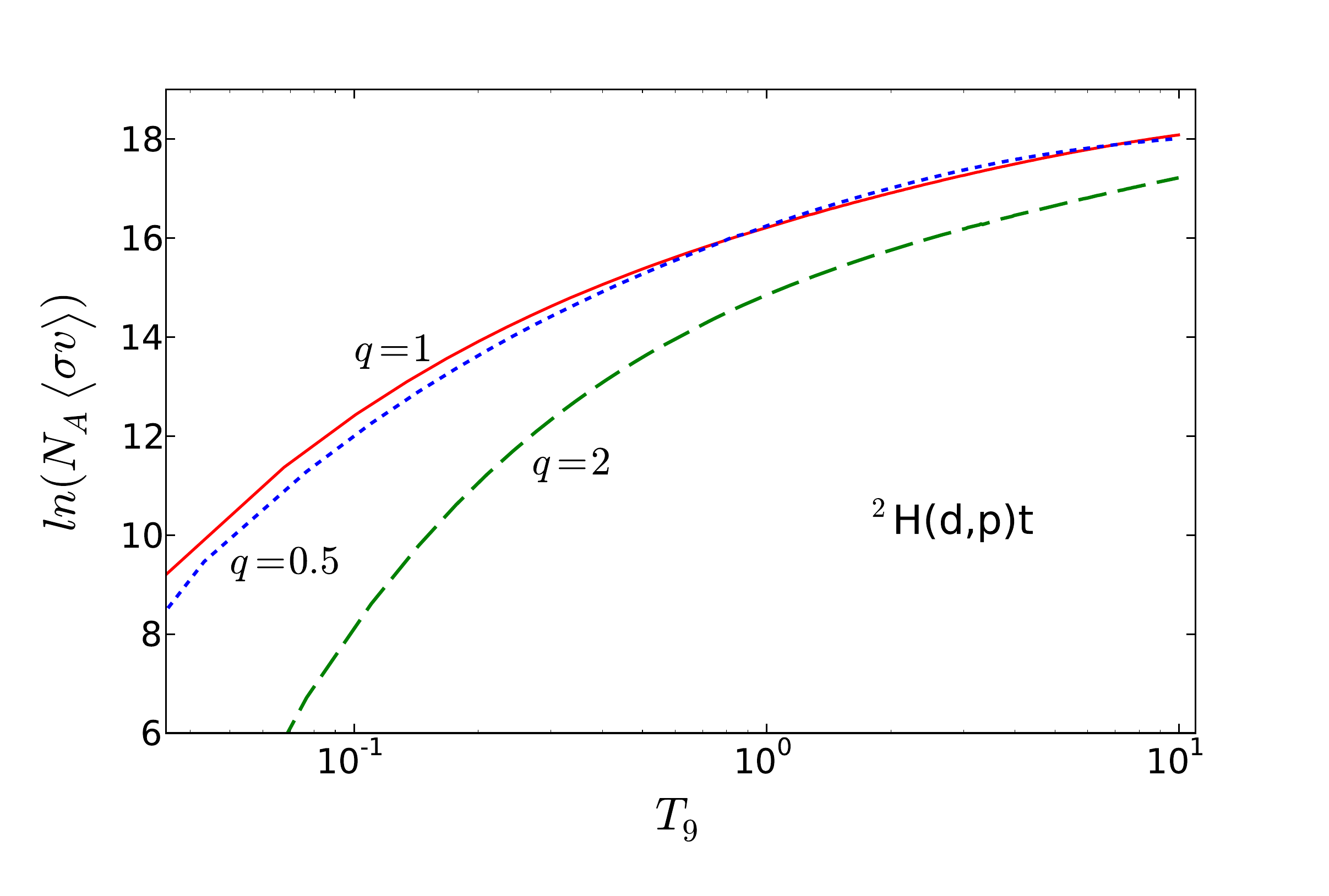}
\caption{Reaction rates for $^2$H(d,p)$^3$H as a function of $T_9$ for various values of $q$. Rates are depicted as a function of the natural logarithm of $N_A \langle \sigma v \rangle$. Results utilizing are shown for  $q=0.5$ (dotted line) and $q=2$ (dashed line).}\label{rddpt}
\end{figure}
\end{center}

We illustrate in Figure \ref{rddpt} the reaction rates for $^2$H(d,p)$^3$H as a function of $T_9$, considering $q=0.5$, 1, 2.  The solid curve represents the conventional Maxwell-Boltzmann distribution ($q=1$). On the other hand, the dashed and dotted curves correspond to $q=2$ and $q=0.5$ respectively. In both instances, deviations from the Maxwellian rate are evident. For $q>1$, the deviations are notably significant, leading to an overall suppression of the reaction rates, particularly at lower temperatures. This deviation stems from the non-Maxwellian energy cutoff, occurring at $0.086$ MeV for this reaction, thereby hindering numerous reactions from occurring at higher energies. For $q<1$, the similarity in results to the Maxwell-Boltzmann distribution arises from a delicate balance between the suppression of reaction rates at low energies and their augmentation at high energies.

This clarifies why, at $T_9=1$, the relevant energy range for computing the reaction rate is delineated by the solid curve in Figure \ref{fmaxwell}. For $q<1$, the Gamow window $\Delta E$ expands, contributing both to the suppression of reaction rates at low energies, akin to the Maxwell-Boltzmann distribution, and to a corresponding enhancement at higher energies. Consequently, this elucidates the nearly identical outcomes depicted in Figure \ref{rddpt} for $q=1$ and $q<1$. This observation holds true for all charged particle reaction rates, except in cases where the S-factor exhibits a pronounced energy dependency at, or around, $E=E_0$. However, such behavior is absent in the most crucial charged-induced reactions during BBN. 

\begin{table}[h]
\caption{Theoretical predictions and observed abundances for primordial light elements}\label{tab1}%
\begin{tabular}{@{}llll@{}}
\toprule
   & $q=1$   & $q=1.069$ - $1.082$& Observation \\
\midrule
$^4$He    & 0.2476   & 0.2470& $0.2453\pm 0.0034$ \cite{Aver21}  \\
D/H($\times10^{-5}$)    & 2.57   & 3.14 - 3.25 & $2.527 \pm 0.030$ \cite{Coo18}  \\
$^3$He/H($\times10^{-5}$)    & 1.04   & 1.46 - 1.50 & $<1.1\pm 0.2$ \cite{BRB02}  \\
$^7$Li/H($\times10^{-10}$)    & 5.23   & 1.62 - 1.90 & $1.58\pm 0.31$ \cite{Sb10}  \\
\botrule
\end{tabular}
\label{tab1}
\end{table}

\begin{center}
\begin{figure}[t]
\centering
\includegraphics[width=0.6\textwidth]{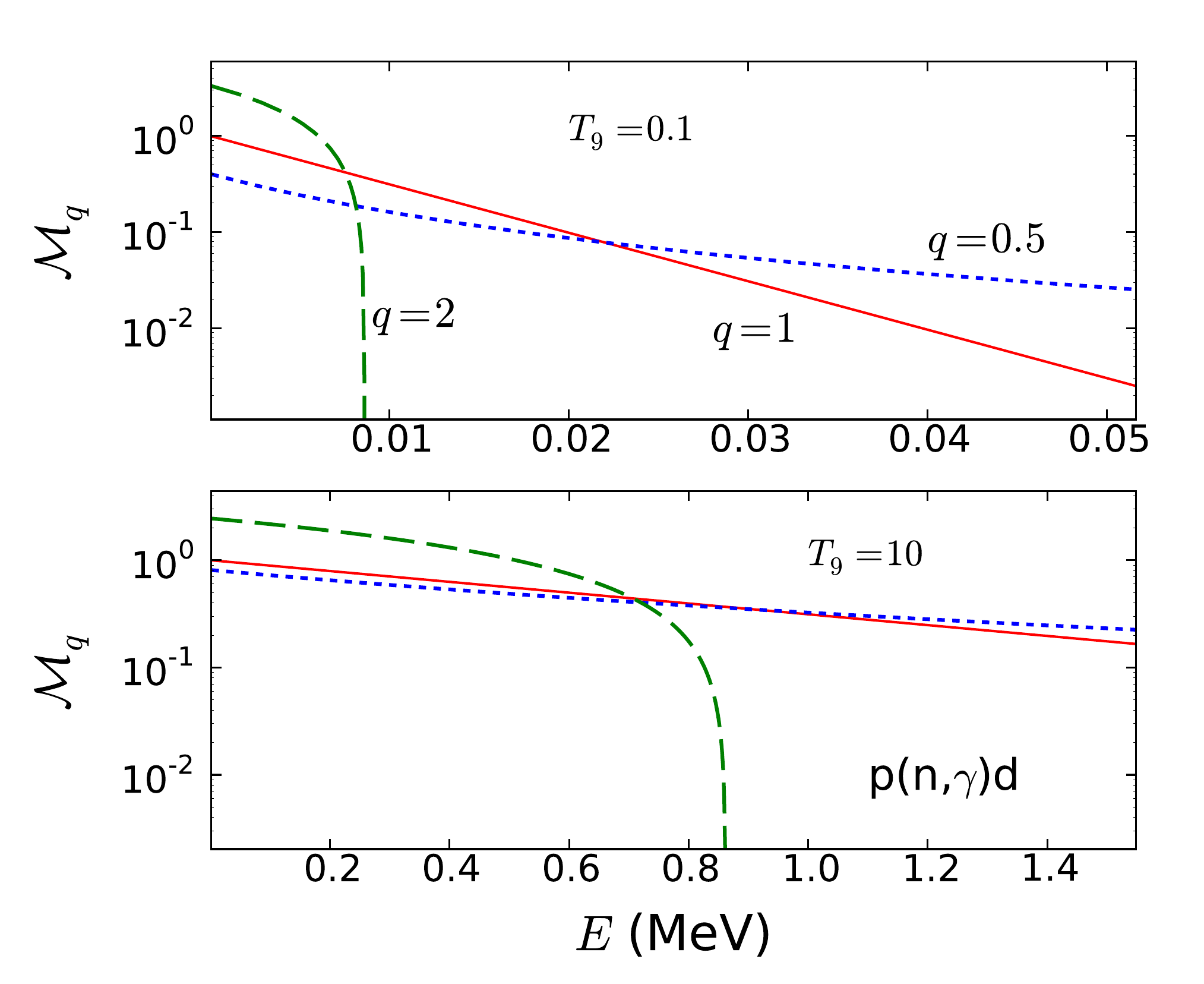}
\caption{${\cal M}_{q}(E,T)$ distributions for protons and neutrons of importance for the p(n,$\gamma$)d reaction for $T_9=0.1$ (upper panel) and $T_9=10$ (lower panel). The solid line, corresponding to $q=1$, represents the Boltzmann distribution. The  non-extensive distributions are shown for for $q=0.5$ (dotted line) and $q=2$ (dashed line). }\label{png}
\end{figure}
\end{center}

In Figure \ref{png}, we depict the distributions of the kinetic energy of nucleons relevant to the p(n,$\gamma$)d reaction at $T_9=0.1$ (upper panel) and $T_9=10$ (lower panel). $q=1$ is represented by the solid line,  with the dotted line for $q=0.5$ (dotted line) dashed line for $q=2$. Akin to the case of charged particles,  higher kinetic energies are more probable with $q<1$, whereas high energies are less accessible with $q>1$ and a cutoff is reached. A notable difference is the absence of the Coulomb barrier, leading to a lack of suppression of the reaction rates at low energies. When $T_9$ increases, the MB and non-MB distributions approach each other. 

\subsection{Backward Reactions}
One usually defines the $1 + 2 \rightarrow 3 + 4$ reaction with positive $Q$-value as the forward reaction and  $3 + 4 \rightarrow 1 + 2$  with a negative $Q$ value as the reverse one. The ratio between the reverse and forward rates is  proportional to $\exp(-Q/kT)$. The non-extensive velocity distribution for reverse reactions is modified to
\begin{eqnarray}
{\cal M}_{q}(E,T)&=&c\times {\cal A}(q,T)\left[1-(q-1)\frac{E+Q}{kT}\right]^{\frac{1}{q-1}}e^{-b/\sqrt{{E}}} ,
\label{MqETr}\end{eqnarray}
where, from the detailed balance theorem,
\begin{equation}
c = \frac{m_1m_2 (2J_1+1)(2J_2+1)(1+\delta_{34})}{m_3m_4 (2J_3+1)(2J_4+1)(1+\delta_{12})}.
\end{equation}
It was first observed in Ref.~\cite{Hou17} that the inclusion of $Q$-values in the reverse reactions has a large impact on the element abundance calculated in the BBN. 
\begin{center}
\begin{figure}[h]
\centering
\includegraphics[width=0.7\textwidth]{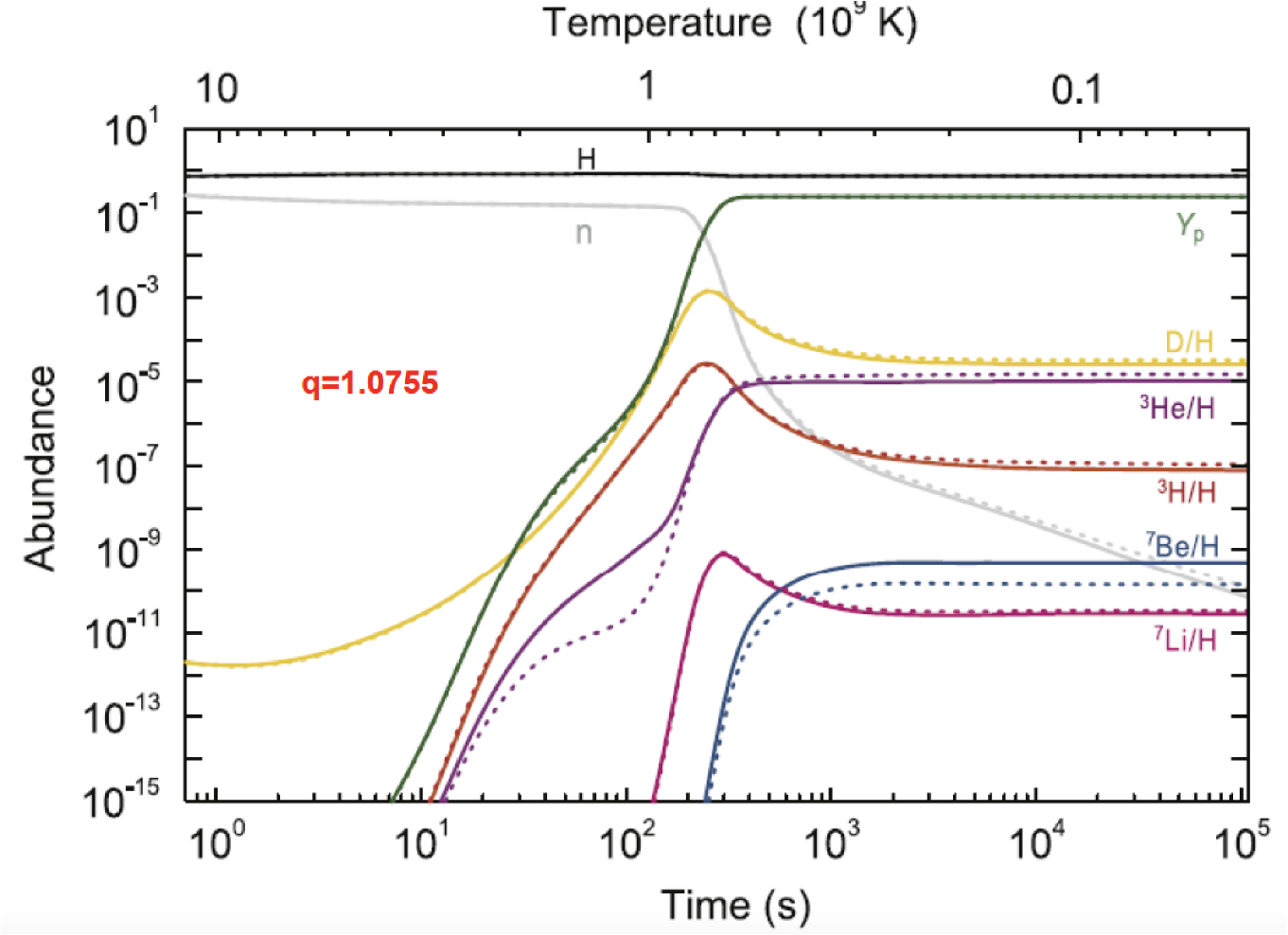}
\caption{Time and temperature evolution of primordial light-element abundances during the BBN era. The solid and dotted lines indicate the results for the classical MB distribution ($q=1$) and the non-extensive distribution ($q=1.0755$), respectively. }\label{bbn2}
\end{figure}
\end{center}

\subsection{Solving the lithium puzzle, creating a deuterium problem}

The primordial abundances are calculated using a standard BBN code and adopting the  cosmological baryon-to-photon ratio $\eta = (6.104 \pm 0.055) \times 10^{-10}$ \cite{Agh2020}, and the neutron lifetime of $\tau = (878 \pm 0.3)$ \cite{Gonz21}.  The thermonuclear (forward and reverse) rates for those 17 principal reactions have been determined employing non-extensive statistics, including 11 reactions of primary importance and 6 secondary reactions in the primordial light-element nucleosynthesis. Standard MB rates have been adopted for the remaining reactions, as they play only a minor role during BBN.  Some reaction rates have been updated by using recent investigations~\cite{Hart18,BSM18,Mukh16,BMS16,Lami17,Spar17}.

Table \ref{tab1}  shows that the predicted and observed abundances of D, $^4$He, and $^7$Li are in close agreement  when a non-extensive velocity distribution with $1.069 < q <1.082$ is adopted.  The $^4$He abundance is rather resilient to the variation of $q$ and is in agreement with the data from Ref.  \cite{Aver21}.  The predicted $^3$He abundance for the above range of $q$ agrees at the 1.8$\sigma$ level with the observed abundance \cite{BRB02}. The prediction of $^7$Li is greatly improved and agrees with the observation \cite{Sb10}.  Thus,  the cosmological lithium problem can be solved with the non-extensive statistics. However, the deuterium abundance now deviates appreciably from the value one would obtain with the standard BG statistics ($q=1$).

The agreement between the $^7$Li abundance and observational data, when considering non-extensive statistics, stems from the reduced production of $^7$Li and radioactive $^7$Be (which decays into $^7$Li) when $q>1$. The production of these elements primarily occurs through radiative capture reactions, namely $^3$H($\alpha,\gamma$)$^7$Li and $^3$He($\alpha,\gamma$)$^7$Be. However, when $q>1$, the alpha-capture rates of these reactions decrease due to the decreased availability of high-energy nucleons compared to the standard MB ($q=1$) distribution. Conversely, the reverse photodisintegration rates remain unaffected by $q$ due to the adoption of Planck’s radiation law for photon energy density. Consequently, the overall production of $^7$Li and $^7$Be decreases, resulting in a better match between predicted and observed primordial abundances. Figure \ref{bbn2}, illustrates the temporal and thermal evolution of primordial abundances during BBN calculated using both MB and non-extensive distributions (with an average value of $q=1.0755$). It demonstrates a significant reduction in the predicted abundance of $^7$Be (which eventually decays into $^7$Li) compared to $q=1$, ultimately offering a solution to the $^7$Li problem within this model.

Deuterium is a very fragile isotope that is surely destroyed after BBN throughout stellar evolution. The tension with the predicted abundances using $q=1.0755$ is explained because as seen from Figure \ref{rddpt}, the reaction rates for $^2$H(d,p)$^3$H tend to decrease for any value of $q>1$. Therefore, there is more deuterium left over from BBN. The fact that we find excellent agreement between observed and  predicted primordial abundances of $^7$Li  in the interval $1.069<q<1.082$ is intriguing. It shows that the solution of the so-called lithium puzzle might be due to a small change in the physics during the BBN. But, it seems that the solution proposed in Ref. \cite{Hou17}, while solving the lithium puzzle leads to a discrepancy with the observed value of $2.527 \pm 0.030 \ (\times 10^{-5})$ for the  D/H abundance  \cite{Coo18}. With the standard MB statistics, the BBN deuterium abundance compares much better with the observations \cite{Pitr21}.

\section{Conclusions}
 The concentrations of the four light elements, namely D, $^3$He, $^4$He, and $^7$Li, serve as robust constraints for primordial abundance assessment. Across all these elements, non-extensive statistics with the non-extensive parameter $q$ slightly above the unity manifest greater agreement with experimental findings. A chi-square fitting against observed elemental abundances suggests that the non-extensive parameter  $q = 1.0755$~\cite{Hou17} best describe observations. 

Presently, $^3$He measurements are only viable within our Galaxy's interstellar medium, limiting assessments at low metallicity crucial for fair primordial element comparisons. Consequently, determining the primordial $^3$He abundance remains unreliable.  Our analysis has excluded potential alterations in n/p conversion rates induced by non-extensive statistics, alongside expected electron distribution changes. These adjustments would influence freezeout temperatures, consequently impacting $^4$He abundance.

Using a non-extensive MB statistics to replicate observed light element abundances without compromising other successful BBN predictions appears unfeasible if one considers its discrepancy with the deuterium observations. Should confirmation of a non-Maxwellian distribution stemming from Tsallis non-extensive statistics emerge, significant revisions to our comprehension of cosmic universe evolution would be imperative.

\bmhead{Acknowledgements}

C.A.B. acknowledges support by the U.S. DOE Grant DE-FG02-08ER41533 and the Helmholtz Research Academy Hesse. Shubhchintak acknowledges SERB, DST, India for a Ramanujan Fellowship (RJF/2021/000176).

\end{document}